\documentstyle[twocolumn,epsf,aps,amsmath,amssymb]{revtex}

\newcommand{\Wf}{ {\mathbf W} } 
\newcommand{\pif}{ \boldsymbol{\pi}}
\begin{document}

\twocolumn[\hsize\textwidth\columnwidth\hsize\csname@twocolumnfalse\endcsname

\title{
A New Stochastic Strategy for the Minority Game
}   
\author{G. Reents, R. Metzler, and W. Kinzel}
\address{Institut f\"ur Theoretische Physik, Universit\"at W\"urzburg,
Am Hubland, D-97074 W\"urzburg, Germany} 
\maketitle

\begin{center} July 21, 2000 \end{center}

\begin{abstract}
We present a variant of the Minority Game in 
which players who were successful in the previous timestep
stay with their decision, while the losers change 
their decision with a probability $p$. Analytical 
results for different regimes of $p$ and the number
of players $N$ are given and connections to existing
models are discussed. It is shown that for $p \propto 1/N$ the
average loss $\sigma^2$ is of the order of $1$ and does not increase 
with $N$ as for other known strategies.

\vspace{3mm}
\vspace{5mm}
\end{abstract} ]

\section{Introduction}
Game theory decribes situations in which players
must make decisions, i.e. choose between different
alternatives, and receive payoffs according to their 
and the other players' choices. The question how 
players decide on a strategy, i.e. how they find out
what to do if they do not possess full information
on the strategies of the other players, was addressed
in Ref. \cite{Arthur:ElFarol}. There it was suggested
that each player has a number of models that prescribe
an action for a given state of the player's world,
for example, for a given game history. The model 
that has proven most successful so far is actually
used by the player.

This approach was applied in the Minority Game
introduced and studied in 
\cite{Challet:Emerg.,Challet:Phase,Marsili:Exact}.
The rules for this game and its variations 
are as follows:  
\begin{itemize}
\item There is an odd number $N$ of players.
\item At each time step $t$ each player $i$ makes
a decision $\sigma_i(t) \in \{+1, -1\}$, 
the majority is determined, 
$S(t) = \mbox{sign}\left(\sum_{i=1}^N \sigma_i(t)\right)$,
and those players who are in the minority, $\sigma_i(t) = -S(t)$, win,
the others lose.
\item A measure of global loss is 
\begin{equation}
\sigma^2 = \left \langle \left (\sum_{i=1}^N \sigma_i(t) \right )^2 
\right \rangle_t. 
\end{equation}
Random guessing leads to $\sigma^2 = N$\,.
\item The only information accessible to players is the 
history of the majority ($S(t-M),\dots, S(t)$). 
In many cases, the history can be replaced
by a random sequence without essentially affecting the results 
\cite{Cavagna:Memory}.
\item Accordingly, no contracts between players are allowed.
\end{itemize}

In the original Minority Game, each player has a small
number of randomly picked decision tables that prescribe
an action for each possible history. Those tables receive
points according to how well they have predicted the 
best action in the course of the game, and the best 
table is used to actually make the decision.

Other publications studied variants in which the
agents used neural networks to make their decisions 
\cite{Kinzel:Interact},
or in which each agent has a probability that determines
whether he chooses the action that was successful in the
last step or its opposite \cite{Johnson:Trader,Lo:Theory}.

\section{The Model}
In this paper, we introduce a very simple prescription
for the agents that still is a reasonable way of 
behaving in the absence of detailed information.
It is in some ways related to Johnson's model 
\cite{Johnson:Trader,Lo:Theory}, but different in decisive details. 
The model is this:
\begin{itemize}
\item If an agent $i$ is successful in a given turn, he
will make the same decision the next turn:
$\sigma_i(t+1) = \sigma_i(t)$ . After all,
there's no reason to change anything. 
\item Otherwise, the agent will change his output
with a probability $p\,$:
\linebreak[2] 
${\sf prob}\left(\sigma_i(t+1) = -\sigma_i(t) \right) = p$.
The agent is reluctant to give up his position,
but eventually, something must change.
\end{itemize}
This is evidently a stochastic one-step process 
and can be handled well with the tools for Markov 
processes. We therefore introduce variables to 
describe an ensemble of games. 

Instead of using the whole set $\{\sigma_i(t)\}_{i=1}^N$ of time
dependent random variables we consider the stochastic process 
\begin{equation}
K(t) = \frac{1}{2}\sum_i \sigma_i(t)\,.
\end{equation}
The possible values $k$ that $K(t)$ can take are
half-integer and run from $-N/2$ to $N/2$ in steps of $1$. Then,
the probabilities
\[ \pi_k(t) = {\sf prob}\left(K(t) = k\right) \]
together with the transition probabilities
\[ W_{k\ell}={\sf prob}(K(t+1)=k \ | \ K(t)=\ell) \] 
are the basic quantities to describe the system. To shorten notation we
consider the probabilities $\pi_k(t)$ as components of the state vector 
$\pif(t) = (\pi_{-N/2}(t),\ldots,\pi_{N/2}(t))^T$. 
The number of players in the majority at time $t$ is $N/2
+|K(t)|$. Since the individual players perform independent Bernoulli
trials, the transition probability $W_{k\ell} = W(\ell \to k)$
from a state with $K(t)=\ell$ to $K(t+1)=k$ is given by the binomial 
distribution  
\begin{eqnarray}  \label{mat-binom}
W_{k\ell} &=& 
  \binom{\frac{N}{2} + \ell }{\ell - k} p^{\ell - k} (1-p)^{\frac{N}{2}+k} 
  \mbox{\ \  for\ } \ell > 0\,, \nonumber \\
W_{k\ell} &=& 
  \binom{\frac{N}{2}-\ell}{k-\ell} p^{k-\ell} (1-p)^{\frac{N}{2}-k}  
\mbox{\ \  for\ } \ell < 0\,. 
\end{eqnarray}
It is understood that $\binom{\frac{N}{2}+|\ell|}{m}=0$ for $m<0$.

This stochastic process may be considered a random walk in one
dimension, where steps of arbitrary size with probability (\ref{mat-binom})
are allowed only in the direction of the origin.

Given the initial state $\pif(0)$, the state $\pif(t)$
is updated at each time step by multiplying it by the transition matrix
$\Wf$:
\begin{equation} \label{dynamics}
\pif(t+1)= \Wf \pif(t)\,. 
\end{equation}
The mathematical
theory dealing with this kind of problems is that of Markov chains with 
stationary transition probabilities \cite{Feller}. Since 
$(\Wf^2)_{k\ell} > 0$, the chain is irreducible as well as
ergodic \cite{Gantmacher}, which implies that irrespective of the initial 
distribution the state $\pif(t)$ converges for $t \to \infty$ 
to a unique stationary state $\pif(\infty) \equiv 
\pif^s$. In view of Eq.\,(\ref{dynamics})\ 
$\pif^s$ corresponds to an eigenvector 
of $\mathbf{W}$ with eigenvalue $1\,$:
\begin{equation} \label{eigenvector}
\Wf\,\pif^s = \pif^s \ \ \ \mbox{and}
\ \ \ \sum_k\pi^s_k = 1.
\end{equation}
The properties of this eigenvector, which by the stated normalization
condition becomes unique, are our main interest.

The problem can be simplified by exploiting the
symmetry $W_{-k,-\ell}=W_{k\ell}$, which implies the symmetry 
$\pi^s_{-k} = \pi^s_k$ of the stationary state. Reformulating the
eigenvalue problem for the independent components of $\pif^s$,
the eigenvector can be calculated
numerically up to $N\approx 1200$ in reasonable time 
with standard linear algebra packages. 

\section{Solution for small probabilities}
A closer look reveals that as $N\rightarrow\infty$,
there are two scaling regimes for $\sigma^2$, depending
on how $p$ depends on $N$. We will first consider
$p= x/(N/2)$, where $x$ is constant and much smaller 
than $N$. As $N$ is increased, the number of players
that switch sides every turn stays constant to 
first order: since the majority is approximately $N/2$,
on the average $x$ agents will change their opinion.

\begin{center}
\begin{figure}
  \epsfxsize= 1.0\columnwidth
  \epsffile{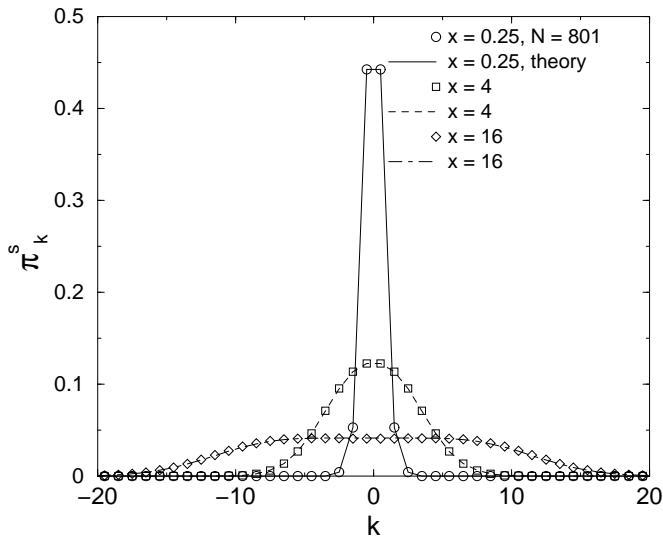}
  \caption{Stationary solution $\pi_k^s$ for $p = 2\, x/N$. The 
    numerical solution for $N=801$ (symbols) is in good agreement with
the analytical solution for $N\rightarrow \infty$.} 
  \label{STO-gammfig}
\end{figure}
\end{center}

In this case the matrix elements $W_{k\ell}$ can be approximated 
by Poisson probabilities \cite{Feller}:
\begin{eqnarray} \label{poisson}
W_{k\ell} &\rightarrow& W^P_{k\ell} = e^{-x} \frac{x^{\ell-k}}{(\ell-k)!} 
   \mbox{\ \ for } \ell > 0, \nonumber \\
W_{k\ell} &\rightarrow& W^P_{k\ell} = e^{-x} \frac{x^{k-\ell}}{(k-\ell)!} 
   \mbox{\ \ for } \ell < 0,
\end{eqnarray} 
where, again, $1/m!$ for negative $m$ has to be interpreted as zero.
In the limit $N\rightarrow \infty$ we are thus looking for an infinite 
component vector $\pif^s$ satisfying the eigenvalue 
equation together with the proper normalization:
\begin{equation} \label{pi-stat}
\Wf^P\,\pif^s = \pif^s \ \ \ \mbox{and}
\ \ \ \sum_k\pi^s_k = 1.
\end{equation}
Making use of (\ref{poisson}) and (\ref{pi-stat}) we were able to derive 
equations for the moments of the stationary distribution:
\begin{eqnarray} \label{moments}
\left\langle |k|-\frac{1}{2}\right\rangle &=& 
\frac{x}{2}\,, \nonumber \\  
\left\langle\left(|k|-\frac{1}{2}\right)
\left(|k|-\frac{3}{2}\right)\right\rangle &=& \frac{x^2}{3}\,, \\
\left\langle\left(|k|-\frac{1}{2}\right)
\left(|k|-\frac{3}{2}\right)
\left(|k|-\frac{5}{2}\right)\right\rangle &=& \frac{x^3}{4}\,,\nonumber \\
& \mbox{etc.} &\nonumber
\end{eqnarray}   
These in turn determine the characteristic function of $\pi^s_k$, 
and a Fourier transform finally leads to 
\begin{equation} \label{pi-stat-sol} 
\pi^s_k = \frac{1}{2\,(|k|-\frac{1}{2})!}\sum_{j=0}^\infty
\frac{(-1)^j\,x^{j+|k|-\frac{1}{2}}}{j!\,(j+|k|+\frac{1}{2})}\,.
\end{equation}
It has been proven that (\ref{pi-stat-sol}) indeed satisfies the
eigenvalue equation (\ref{pi-stat}) \cite{Horn}.
Note that $\pi^s_k$ can be expressed by the incomplete gamma
function:
\begin{equation}
\pi^s_k = \frac{\gamma (|k|+\frac{1}{2},x)}{2\,x\,(|k|-\frac{1}{2})!}\,.
\end{equation}
A comparison with numerically determined eigenvectors of the 
matrix (\ref{mat-binom}) for $N=801$ gives excellent agreement, as seen 
in Fig.\,\ref{STO-gammfig}. 
The distribution is roughly flat for
small $|k|$, has a turning point near $|k| = x$ and falls off 
exponentially with $k$ for larger values of $|k|$. 
From (\ref{pi-stat-sol}), the variance $\sigma^2 = 
\left\langle (2\,k)^2\right\rangle$  can be
calculated:
\begin{equation}
\sigma^2 = 1 + 4\,x + \frac{4}{3}\,x^2. \label{STO-smx}
\end{equation}
For small $x$, this approaches the optimal 
value $\sigma^2=1$ that occurs if the majority is always as narrow as 
possible, but even for larger $x$, $\sigma^2$ does not increase with $N$.

\vspace{-10mm}
\section{Solution for large probabilities}
The other scaling regime assumes that $p$ is of order one
and  $p\,N \gg 1$. To handle this regime, 
we will use a rescaled (continuous) coordinate 
$\kappa = k/N = \sum_i \sigma_i /(2N)$, the range of which is 
$-1/2\leq \kappa \leq 1/2$.
Multiplied by $N$, the stationary
state $\pi^s_k$ for large $N$ turns into a probability density
function $\pi^s(\kappa)$, and the matrix $W_{k\ell}$ becomes an 
integral kernel 
$W(\kappa,\lambda)$, hence (\ref{eigenvector}) is transformed into an 
integral equation: 
\begin{equation} \label{int-eigenvalue}
\pi^s(\kappa) = \int W(\kappa,\lambda)\,\pi^s(\lambda)\,d\lambda\, \ \ \ 
\mbox{and} \ \ \ \int \pi^s(\kappa)\,d\kappa = 1\,.
\end{equation}
\begin{figure}
  \epsfxsize= 1.0\columnwidth
  \epsffile{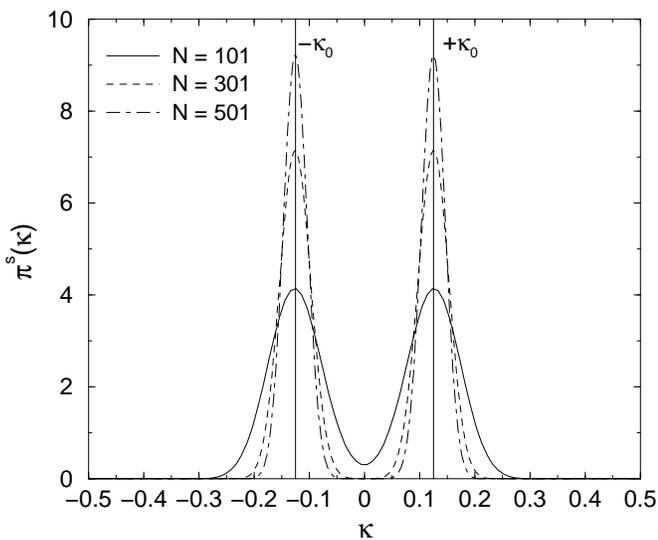}
  \caption{Stationary solution $\pi^s(\kappa)$ for $p = 0.4$. 
With increasing $N$, the width of the peaks becomes
narrower.} 
  \label{STO-p04}
\end{figure}
Numerical calculations show that the eigenvector $\pi^s(\kappa)$ 
takes the shape of two Gaussian peaks centered at symmetrical distances
$\pm \kappa_0$ from the origin (see Fig. \ref{STO-p04}).

The physical interpretation is that the majority switches
from one side to the other in every time step. Since approximately 
$(\kappa_0+1/2)p\,N$ agents switch sides every turn and the 
distance between the two peaks amounts to a number of 
$2\,\kappa_0\,N$ agents, we get $\kappa_0 = p/(4-2p)$. 

This reasoning can be made more precise, and also the width of the 
peaks for large but finite $N$ can be calculated by the following argument:
The well known normal approximation for the binomial coefficients in
(\ref{mat-binom}) leads to
\begin{eqnarray} \label{int-kernel}
W(\kappa,\lambda) = N\,W_{k\ell}  & \approx &
\frac{1}{\sqrt{2\,\pi}\,s(\lambda)} \exp\left[-\frac{1}{2}\frac{(\kappa
 - f(\lambda))^2}{s^2(\lambda)}\right]\,,   \nonumber \\ 
\mbox{where} \ \  
f(\lambda) & = & (1-p)\,\lambda - {\rm sign}(\lambda)\frac{p}{2}
 \nonumber \\
\mbox{and} \ \ 
s^2(\lambda) & = & \frac{p\,(1-p)\,(\frac{1}{2}+|\lambda|)}{N}\,. 
\end{eqnarray}
A double gaussian of the form
\begin{equation} \label{ansatz}
\pi^s(\kappa) = \frac{1}{2}\frac{1}{\sqrt{2\,\pi}\,b}
\left[\exp\left(\frac{(\kappa+\kappa_0)^2}{2\,b^2}\right) + 
      \exp\left(\frac{(\kappa-\kappa_0)^2}{2\,b^2}\right) \right]
\end{equation}
is transformed by the integral kernel (\ref{int-kernel}) into a double 
peak of the same type if in the integral equation we approximate the 
variance $s^2(\lambda)$ of
(\ref{int-kernel}) by $s^2(\pm\kappa_0)$ and if the assumption 
$b^2 \ll \kappa_0^2$ is justified. It means that the peaks are well
separated and that the integral can be extended from $-\infty$ to $\infty$.
By requiring $\pi^s(\kappa)$ from (\ref{ansatz}) to satisfy the
eigenvalue equation (\ref{int-eigenvalue})  we get
\begin{equation}
\kappa_0 = \frac{p}{2(2-p)} \mbox{\ \ and \ }b^2=\frac{1-p}{(2-p)^2 N}.
\label{kappa-width}
\end{equation} 

The result for $\kappa_0$ confirms the simple argument given above, 
whereas the term for $b^2$ is slightly surprising: it does not
depend on $p$ in the leading order, i.e. it is not simply
the number of players who switch sides. Eq. (\ref{kappa-width})
also allows to check whether the assumptions made for its 
derivation are true for a given $p$ and $N$. For example, 
for $p = x/N$, $\kappa_0^2/b^2 \rightarrow 0$ for $N\rightarrow \infty$
according to (\ref{kappa-width}), so one cannot expect 
the formation of double peaks in this limit. The crossover from
single-peak to double-peak distribution occurs for $p \propto 1/\sqrt{N}$. 

\begin{figure}
  \epsfxsize= 1.0\columnwidth
  \epsffile{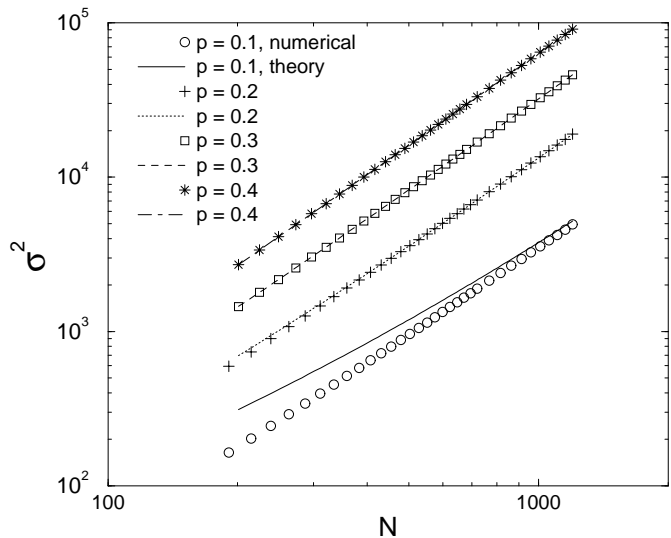}
  \caption{$\sigma^2$ for several values of $p$ and $N$,
compared to predictions by Eq. (\ref{STO-meansm}).} 
  \label{STO-pfix}
\end{figure}

It is now easy to integrate over the probability distribution
to get an expression for $\sigma^2$:
\begin{equation}
\sigma^2 = \frac{N}{(2-p)^2}(Np^2 + 4(1-p)). \label{STO-meansm}
\end{equation}
This holds well if the condition $\kappa_0 \gg b$ is fulfilled,
i.e. for sufficiently large $p$ and $N$, as seen in Fig.
\ref{STO-pfix}. 

\section{Concluding remarks}
The presented strategy can be related to the decision tables
of Challet and Zhang's Minority Game as follows: if every 
player keeps only one decision table with entries for all 
possible histories and changes the entries individually
with a probablility of $p$ if he loses in a given time step,
the mean result will we exactly the same as for the presented
one-step memory. A similar argument was given for Johnson's
variant in \cite{Burgos:Selforg.}. The memory size, which 
determines the number of entries in the tables, is completely
irrelevant for the average loss of each player, but does influence
the time series of minority decisions generated by the system.

In summary, we have found an analytic solution of a stochastic
strategy for the Minority Game. Although this strategy is very simple,
it yields an average loss of order one even  in the limit 
of infinitely many agents. Questions that will be discussed
in future publications include the dynamics and relaxation 
time of the system, interactions with players using other
strategies and individual probabilities for each player. 

\section{Acknowledgement}
R. M. and W. K. acknowledge financal support by the German-Israeli
Foundation. We would like to thank Christian Horn, Andreas Engel
and Ido Kanter for helpful discussions.

\end{document}